# The Development of Coplanar CZT Strip Detectors for Gamma-Ray Astronomy

M. L. McConnell†, L.-A. Hamel‡, J. R. Macri†, M. McClish†, and J. M. Ryan†

†*Space Science Center, University of New Hampshire, Durham, NH 03824*
‡*Physics Department, University of Montreal, Montreal, Quebec, Canada*

**Abstract.** The development of CdZnTe detectors with an orthogpnal coplanar anode structure has important implications for future astrophysical instrumentation. As electron-only devices, like pixel detectors, coplanar anode strip detectors can be fabricated in the thickness required to be effective for photons with energies in excess of 500 keV. Unlike conventional double-sided strip detectors, the coplanar anode strip detectors require segmented contacts and signal processing electronics on only one surface. This facilitates the fabrication of closely-packed large area arrays that will be required for the next generation of coded aperture imagers. These detectors provide both very good energy resolution and sub-millimeter spatial resolution (in all three spatial dimensions) with far fewer electronic channels than are required for pixel detectors. Here we summarize results obtained from prototype detectors having a thickness of 5 mm and outline a concept for large area applications.

## INTRODUCTION

Cadmium Zinc Telluride (CZT) detectors are well suited for fabrication of large area, high performance X-ray and γ-ray imaging spectrometers. They have the desirable properties of high stopping power, low thermal noise, room-temperature operation, excellent energy resolution and unsurpassed spatial resolution. With current technologies, high performance CZT imaging spectrometers can be adapted for use in a variety of imaging techniques to image photons at energies up to a few hundred keV. In the future, as the quality of CZT material improves to the point where the use of thicker detector substrates become feasible, the useful energy range of CZT detectors will likely be extended into the MeV energy range.

The spectroscopic value of CZT is exemplified by its use in NASA's upcoming SWIFT mission [1]. Related CdTe detectors are also being deployed on the INTEGRAL mission of ESA [2]. Neither of these applications take advantage of the properties of CZT that allow for the determination of a photon interaction site *within* a detector. Instead, small detector elements serve to isolate the location of the photon interaction site to within several millimeters. Outfitting large areas with numerous small detector elements so that the improved (sub-mm) spatial resolution could be achieved would be prohibitively complex and expensive. The only practical solution would be to employ large-area position-sensitive detectors, drastically reducing the number of detectors and associated electronic channels.

An image plane fabricated using closely-packed arrays of CZT detector modules is a candidate for the central detector of several proposed hard X-ray (30-600 keV) coded aperture telescopes, including HEXIS [3], AXGAM [4], EXIST [5], and MARGIE [6,7]. CZT strip detectors are also under investigation for use as calorimeter detectors in some Compton telescope designs for the 1-100 MeV energy range (e.g., TIGRE [8]). These designs will require image planes with areas exceeding 1000 $cm^2$ and spatial resolutions on the order of 1mm or better.

## THE ORTHOGONAL COPLANAR ANODE STRIP DETECTOR

Good efficiency, energy resolution and position resolution have been demonstrated in pixellated CZT detectors up to 10 mm thick [9]. These detectors are *electron-only* devices that avoid the deleterious consequences of poor hole transport. Unfortunately, the anode pixel contact geometry requires an electronic signal channel for each of the $N^2$ pixels. Strip detectors, on the other hand, provide $N^2$ pixels with only 2N electronic channels, one for each row and one for each column. This is important in space flight applications where channel count greatly affects the complexity of the instrument design. Double-sided CZT strip detectors, if carefully designed, can address many of the limitations associated with poor hole transport [10], but they will still be limited in terms of detector thickness (~3 mm) and, consequently, the effective energy range (<300 keV).

We have been developing a novel CZT detector concept: an *electron-only* device featuring *orthogonal coplanar anode strips* [11-14]. Each row takes the form of N discrete interconnected anode pixels, while each column is a single anode strip. Figure 1 illustrates the design of an 8 × 8 *pixel* orthogonal coplanar anode strip detector. The opposite side has a single uniform cathode electrode. The anode pixel contacts, interconnected in rows, are biased to collect the electron charge carriers. The orthogonal anode strips, surrounding the anode pixel contacts, are biased between the cathode and anode pixel potentials. The strips register signals from the motion of electrons as they migrate to the pixels. Since electrons are much more mobile than holes in CZT, signals from photon interactions at all depths in the detector are detected. Given the published results with pixel detectors [9], we expect our coplanar anode strip approach to be effective in CZT detectors at least 10 mm thick. This will permit thicker, more efficient, CZT imaging planes than are practical with double-sided strip detectors and will extend the effective energy range to >1 MeV. More compact packaging is also possible since all electrical connections for processing are on one side of the detector.

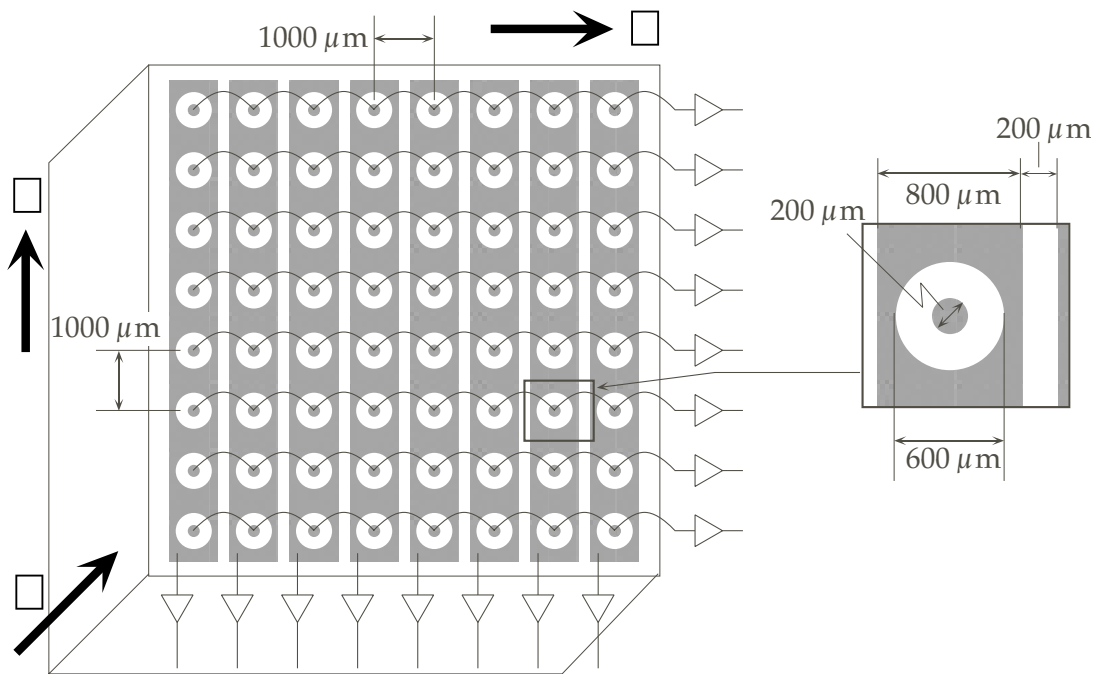

**FIGURE 1.** A schematic diagram showing the layout of the orthogonal coplanar anode design as used in our prototype detectors. Strip columns (X) are read out on the bottom. Pixel rows (Y) are read out on the right. The CZT thickness is 5 mm.

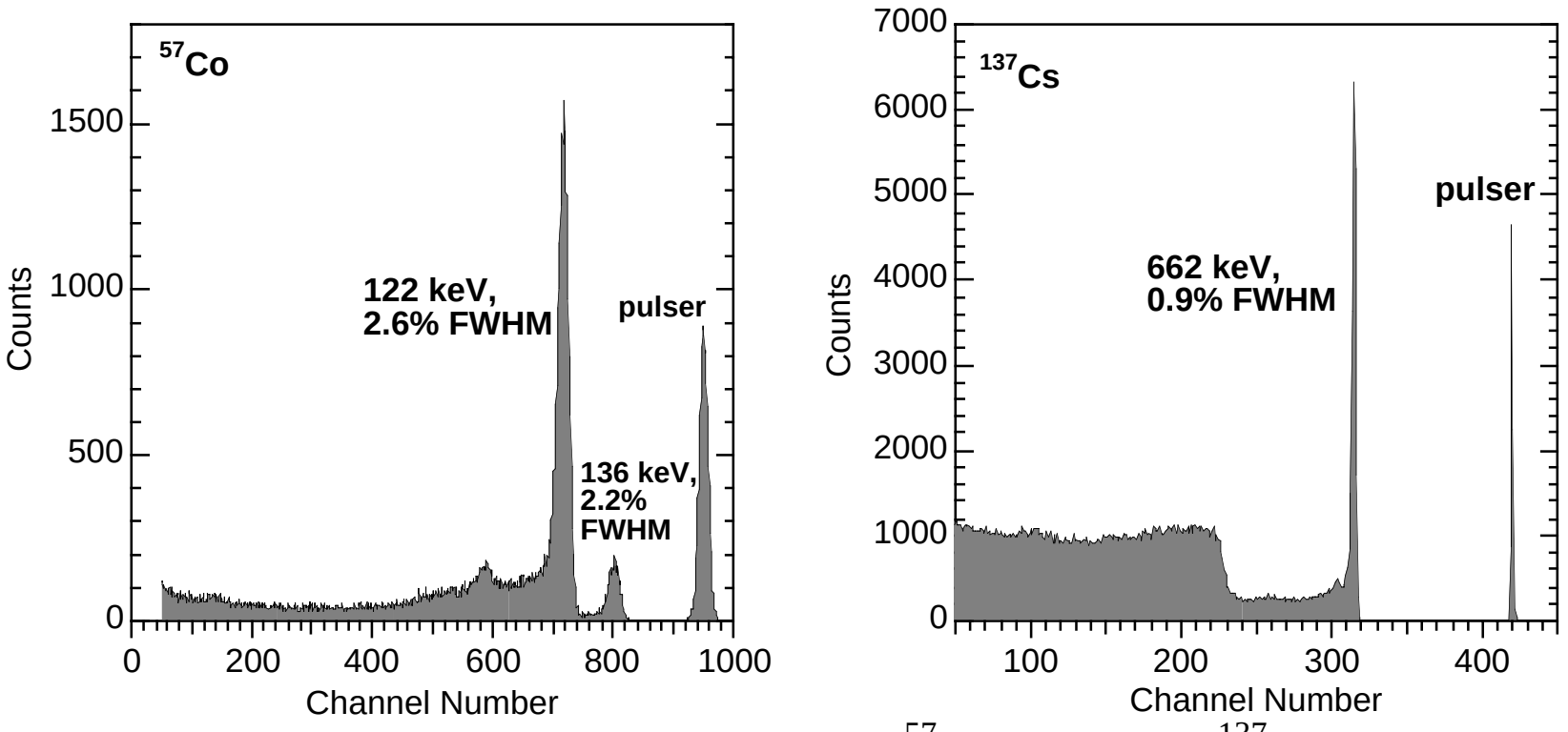

**FIGURE 2.** Single pixel spectra from $^{57}$Co (left), and $^{137}$Cs (right).

# PROTOTYPE PERFORMANCE

We have fabricated several prototype detectors based on the orthogonal coplanar anode strip design. Each prototype module consists of a patterned CZT substrate that has been polymer flip-chip (PFC) bonded to a ceramic (LTCC) substrate or multi chip module (MCM) [15]. Reliable connections to all strips and pixel rows have been achieved with this fabrication approach. The spectroscopic performance of a single pixel element is demonstrated in Figure 2. The test pulse data shown for two of these spectra indicate that electronic noise presently limits the energy resolution and suggests that further improvements may still be possible.

Some level of charge sharing between adjacent strips/pixels is required in order to infer locations from relative signal measurements. Some charge sharing occurs in both directions, although the level of charge sharing in the X-direction is somewhat higher due to the larger surface area of the strips. This results in somewhat better spatial resolution in the X-direction. Alpha particle scans across the surface of the detector have yielded 1σ spatial resolutions of ~100 μm in X and ~300 μm in Y. The interaction depth (Z) can be inferred from the ratio of the amplitudes of the cathode and anode signals. We have demonstrated a 1σ FWHM position resolution of 650 μm in the Z-coordinate [13, 14].

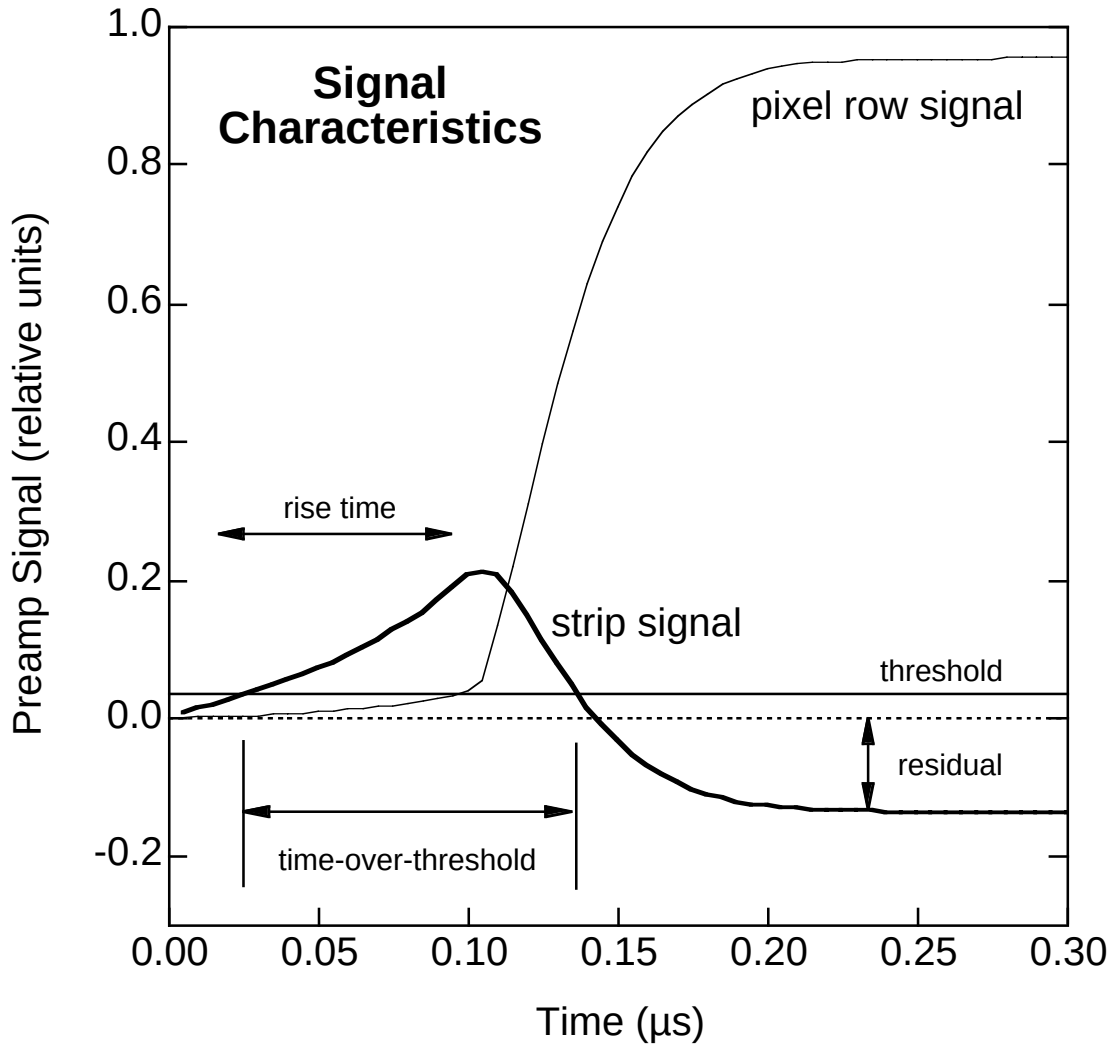

**FIGURE 3**. Signal characteristics associated with the pixel rows and the orthogonal strips. The strip signal parameters are related to the depth of interaction.

While the cathode signal can be used for the depth of interaction measurement, the need to feed the cathode signals from the front surface of the detector to front-end electronics located behind the detector will interfere with the fabrication of closely packed arrays of detector modules. We have therefore been studying approaches that will permit the measurement of the interaction depth using only those signals that are available on the back (anode) side of the detector [16]. These signals are shown in Figure 3.

The pixel signals, rising in only the positive direction, are typical of small-pixel anodes in CZT detectors. The initial slope of the pixel signals is small but increases rapidly when electrons reach the anode region. These signals provide a measure of the energy deposit and identify the Y-coordinate of the photon interaction location. The strip signals identify the X-coordinate of the photon interaction location, but are not used for an energy measurement. The strip signals are bipolar in nature. They have faster initial rises than the pixel signals due to the larger strip areas. They reach a maximum shortly before electron transit time and decrease as the electrons approach the pixel. Of particular interest are three features of the anode strip signal (risetime, time-over-threshold, and residual; see Figure 3) that can be used to measure the depth of interaction, independent of any signal from the cathode. We have been working to design and implement a circuit that can directly measure the interaction depth, using the time-over-threshold of the bipolar strip signal [16].

## FABRICATION CONCEPT FOR LARGE-AREA ARRAYS

The baseline approach for packaging a single detector module is illustrated in Figure 4. Here we envision a module with $16 \times 16$ logical pixels formed from 16 strips and 16 pixel rows, each on a 1 mm pitch. The CZT substrate in this case is roughly $16 \times 16$ mm$^2$.

The PFC bonding process establishes the mechanical and electrical interconnection between the CZT and LTCC substrates. The multi-layer LTCC substrate establishes the interconnection of the anode pixels in rows and the routing of the pixel row and strip signals (total of 32) plus the guard ring contact to flat gold contact pads on the underside of the module. The fact that 256 *pixels* are achieved with only 32 signal processing channels leaves ample space on the underside of the LTCC substrate for the necessary signal processing electronics

Figure 5 shows the concept of an image plane board supporting a $20 \times 20$ CZT module array having 1024 cm$^2$ active area. The image plane is a large-area, mechanically reinforced circuit board that supports an array of closely packed CZT detector modules.

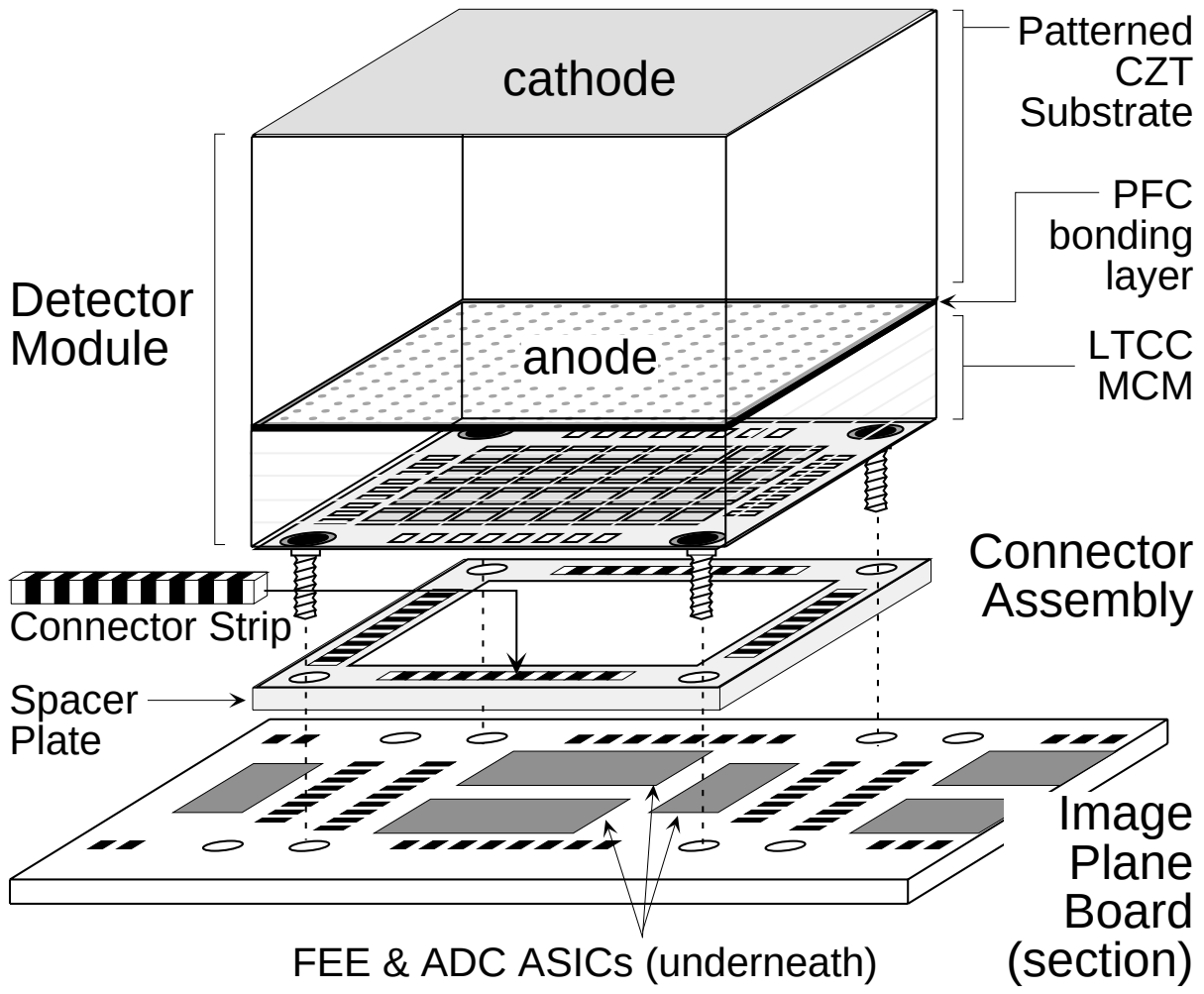

**FIGURE 4**. Concept for a detector module to be used in fabrication of large area arrays.

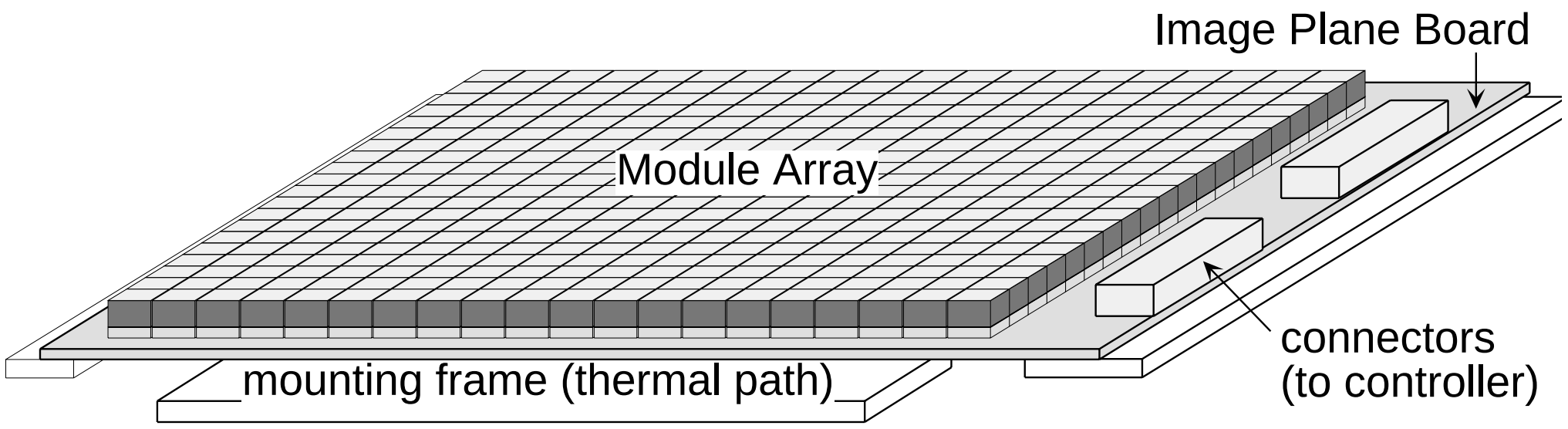

**Figure 5**. Image plane concept. A 20 × 20 array of imaging CZT modules.

The estimated packing fraction may be as high as 94%. The board supports detector bias and the front-end electronics. These electronics are on the underside of the board. The bonding and connector assembly technologies provide good thermal isolation between the image plane board and the CZT. The cathode bias will be provided using a thin flexible circuit located between rows of modules (not shown).

Assuming 2 mW/channel for the pixel and strip FEE ASICs, the total FEE power dissipation for the array shown in Figure 4 would be 26 W. A pixel detector array of the same size would require 205 W. In our design, the heat dissipated on the image plane-board is conducted to the experiment structure via its mounting frame. The electrical interfaces between the image plane and the image-plane controller are digital, thereby permitting flexibility in the location of the controller electronics.

## ACKNOWLEDGMENTS

This work is supported at UNH by grants from NASA's High Energy Astrophysics SR&T program and at UM by the Natural Sciences and Engineering Research Council (NSERC) of Canada.